\documentclass[12pt,fleqn]{article}
\usepackage{amsmath}
\usepackage{amssymb} 
\usepackage{mathtools}
\usepackage{dsfont} 
\usepackage[small]{caption} 
\usepackage{subcaption}
\usepackage{booktabs}
\usepackage{mathrsfs}	
\usepackage{cite} 
\usepackage{cancel}
\usepackage{nicefrac}
\usepackage{pifont} 
\usepackage{geometry}
\usepackage[linkcolor=blue,urlcolor=blue,citecolor=magenta,colorlinks=true]{hyperref}
\usepackage{cleveref} 
\usepackage{braket}
\usepackage{xfrac}
\usepackage{wrapfig}
\usepackage{microtype}
\usepackage[nolist]{acronym}

\makeatletter
\g@addto@macro\bfseries{\boldmath}
\makeatother
\DeclareMathOperator{\re}{Re}
\DeclareMathOperator{\im}{Im}

\DeclareMathOperator{\diag}{diag}

\newcommand{\I}{\mathrm{i}}

\newcommand{\SU}[1]{\ensuremath{\mathrm{SU}(#1)}}

\newcommand{\U}[1]{\ensuremath{\mathrm{U}(#1)}}
\newcommand{\Z}[1]{\ensuremath{\mathds{Z}_{#1}}} 
\newcommand{\ChargeC}{\ensuremath{\mathcal{C}}}

\newcommand{\rep}[2][]{\ensuremath{{\boldsymbol{#2}_{#1}}}}

\allowdisplaybreaks

\UseCollection{xfrac}{plainmath}

\begin{document}

\newcommand\mytitle{Quasi--Eclectic Modular Flavor Symmetries}

\begin{titlepage}
\begin{flushright}
UCI--TR--2021--20%
\\
\end{flushright}

\vspace*{1em}

\begin{center}
{\Large\sffamily\bfseries\mytitle}

\vspace{1em}

\renewcommand*{\thefootnote}{\fnsymbol{footnote}}

\textbf{%
Mu--Chun Chen$^{a,}$\footnote{muchunc@uci.edu}, 
V\'ictor Knapp--P\'erez$^{b,c,}$\footnote{victorknapp@ciencias.unam.mx},
Mario Ramos--Hamud$^{b,}$\footnote{hamud@ciencias.unam.mx},\\
Sa\'ul Ramos--S\'anchez$^{b,}$\footnote{ramos@fisica.unam.mx},
Michael Ratz$^{a,}$\footnote{mratz@uci.edu} 
and Shreya Shukla$^{a,}$\footnote{sshukla4@uci.edu}
}
\\[8mm]
\textit{$^a$\small
~Department of Physics and Astronomy, University of California, Irvine, CA 92697-4575 USA
}
\\[5mm]
\textit{$^b$\small Instituto de F\'isica, Universidad Nacional Aut\'onoma de M\'exico, POB 20-364, Cd.Mx. 01000, M\'exico}
\\[5mm]
\textit{$^c$\small Address from September 2021: Department of Physics and Astronomy, University of California, Irvine, CA 92697-4575 USA}
\end{center}

\vspace*{1cm}

\begin{abstract}
Modular flavor symmetries provide us with a new, promising approach to the
flavor problem. However, in their original formulation the kinetic terms of the
standard model fields do not have a preferred form, thus introducing additional
parameters, which limit the predictive power of this scheme.  In this work, we
introduce the scheme of quasi--eclectic flavor symmetries as a simple fix. These
symmetries are the direct product of a modular and a traditional flavor
symmetry, which are spontaneously broken to a diagonal modular flavor subgroup.
This allows  us to construct a version of Feruglio's model with the K\"ahler
terms under control. At the same time, the starting point is reminiscent of what
one obtains from explicit string models.
\end{abstract}

\end{titlepage}
\renewcommand*{\thefootnote}{\arabic{footnote}}
\setcounter{footnote}{0}

\section{Introduction}

Modular flavor
symmetries~\cite{Feruglio:2017spp,Penedo:2018nmg,Criado:2018thu,Kobayashi:2018scp,Liu:2019khw,Novichkov:2019sqv,Liu:2020msy,Ding:2020zxw} 
are an exciting new approach to the flavor problem. Very simple settings can, in 
principle, provide us with a surprisingly good fit to data while making a comparatively
large number of nontrivial, testable predictions. We would like to refer the reader 
to~\cite{Feruglio:2019ybq} and references therein for more details and models. 

What is the new ingredient of Feruglio's models~\cite{Feruglio:2017spp} which
appears to make the traditional $A_4$ models \cite{Ma:2001dn,Babu:2002dz} even
more compelling? The challenges in traditional models (see 
e.g.~\cite{Ishimori:2010au} for an extended list of examples and references) lie
mainly in the flavon sector. More specifically, one has to align the flavons at
some appropriate values, see e.g.~\cite{Holthausen:2011vd} for a discussion and
further references. However, often flavons naturally settle at
symmetry--enhanced points (see e.g.\ \cite{Kofman:2004yc}), which are typically
not entirely realistic. As a consequence, the traditional flavor models
often require an extended flavon sector, which introduces a number of free
parameters, thus limiting the number of nontrivial predictions. Models with
modular flavor symmetries evade these arguments because the flavons get replaced
by multiplets of modular forms. One then faces the lesser challenge to find, and
eventually justify, appropriate values of the half--period ratio $\tau$ which
the modular forms depend on. The resulting models are very elegant and describe
data surprisingly well~\cite{Feruglio:2019ybq}.

However, there is a price one has to pay. The modular flavor symmetries, which
we will review in some more detail in \Cref{sec:ModularFlavorSymmetries}, are
nonlinearly realized. As a consequence, the K\"ahler potential is not under
control~\cite{Chen:2019ewa}, i.e.\ there is no preferred field basis. This
introduces additional parameters, thus limiting the predictive power of the
construction. On the other hand, in the framework of traditional flavor
symmetries the K\"ahler potential is under control. It is still subject to
possibly important corrections~\cite{Chen:2012ha,Chen:2013aya}, but one has at
least a perturbative expansion in $\varepsilon=\Braket{\xi}/\Lambda$, where
$\Braket{\xi}$ denotes the \ac{VEV} of a so--called flavon and $\Lambda$ is the
cut--off scale.

The purpose of this study is to show that a hybrid scheme allows us to combine
the advantages of both approaches while largely avoiding their limitations. The
simplest models of this hybrid approach have a flavor symmetry of the form
\begin{equation}
\label{eq:flavorGroup}
 G_\mathrm{flavor}=G_\mathrm{traditional}\times G_\mathrm{modular}\;,
\end{equation}
and a flavon $\chi$, which is charged under both $G_\mathrm{traditional}$ and
$G_\mathrm{modular}$. Once $\chi$ acquires a \ac{VEV}, the flavor symmetry
will be broken to its diagonal subgroup,
\begin{equation}
\label{eq:DiagonalBreaking}
 G_\mathrm{flavor}=G_\mathrm{traditional}\times G_\mathrm{modular}
\xrightarrow{~\Braket{\chi}~}G_\mathrm{diagonal}\;.
\end{equation}
Matter fields are assumed to transform under $G_\mathrm{traditional}$, which is
why their K\"ahler potential is under control
\cite{Chen:2012ha,Chen:2013aya,Nilles:2020kgo}. However, after the
breaking~\eqref{eq:DiagonalBreaking}, their couplings will be effectively given
by modular forms.

Our setup is heavily inspired by the scheme of ``eclectic flavor groups''
\cite{Nilles:2020nnc}, which arise naturally in string models
\cite{Nilles:2020kgo,Nilles:2020tdp,Baur:2021mtl} and magnetized toroidal
compactifications~\cite{Ohki:2020bpo}. Generally in top--down models (cf.\ e.g.\
\cite{Kobayashi:2016ovu,Kobayashi:2018rad,Kobayashi:2018bff,Almumin:2021fbk})
one can, at least in principle, compute the K\"ahler potential, but at this
point it is probably also fair to say that this approach has not yet provided us
with completely realistic predictive models. These groups are the result  of
combining nontrivially a traditional and a modular flavor group, such that 
$G_\mathrm{modular}$ is a subgroup of the outer automorphisms of 
$G_\mathrm{traditional}$. Hence, eclectic groups represent a more complex
hybrid  scheme than~\Cref{eq:flavorGroup}, sharing the feature of a controlled 
K\"ahler potential due to $G_\mathrm{traditional}$. The purpose of the present
work is to show how one can, in a bottom--up \ac{EFT} approach, combine modular
flavor symmetries with perturbative control over the K\"ahler potential. We
leave the question of an explicit stringy completion for future work.

\section{Modular and eclectic flavor symmetries}

\subsection{Modular flavor symmetries}
\label{sec:ModularFlavorSymmetries}

The half--period ratio or modulus $\tau$ of a torus does not uniquely
characterize a given torus. Rather, different $\tau$ related by transformations 
in the so--called modular group $\text{PSL}(2,\Z{})$ describe the same torus. 
Under an arbitrary element $\gamma\in\text{PSL}(2,\Z{})$, the modulus and matter
superfields $\Phi_j$ transform as 
\begin{subequations}
\begin{align}\label{eq:modtrafo}
    \tau &\xmapsto{~\gamma~}
	 \gamma\,\tau :=\frac{a\, \tau+ b}{c\, \tau+ d}\;, \\
    \Phi_{j} &\xmapsto{~\gamma~}
	(c\,\tau+d)^{k_j}\,\rho_{\rep{r}_j}(\gamma)	\, \Phi_{j}
	\;,\quad
    \text{where}~
    \gamma := \begin{pmatrix} a & b \\ c & d \end{pmatrix}
\end{align}
\end{subequations}
with $\det\gamma=1$ and $a,b,c,d\in\Z{}$. Further, $k_j$ denotes the so--called
modular weight of the matter superfield $\Phi_j$, which can build an
$r_j$--dimensional representation of some finite  modular group\footnote{We
restrict here to $\Gamma_N$ finite modular groups, but our discussion can be
readily extended to their double cover $\Gamma_N'$ and metaplectic extensions
(cf.\ \cite{Liu:2019khw,Liu:2020msy,Yao:2020zml,Almumin:2021fbk}).}
$\Gamma_N$, $N=2,3,\ldots$. $\rho_{\rep{r}_j}(\gamma)$ corresponds to the
$r_j\times r_j$ matrix representation  of $\gamma$ in the finite modular group.
This transformation of the matter fields indicates that $\Gamma_N$ can be
regarded as a modular flavor symmetry~\cite{Feruglio:2017spp,Criado:2018thu},
which is however nonlinearly realized, as is evident from~\Cref{eq:modtrafo}.
Finally, note that the action of the modular flavor symmetry is accompanied by
$(c\tau+d)^{k_j}$, which is known as automorphy factor.

Note that, as a consequence of~\Cref{eq:modtrafo}, 
\begin{equation}
  (-\I\tau + \I \bar{\tau} )^{k}\xmapsto{~\gamma~} 
  \bigl((c\,\tau+d) (c\,\bar{\tau}+d)\bigr)^{-k} (-\I\tau + \I\bar{\tau} )^{k}
  \;,
\end{equation}
for an arbitrary $k$. This implies that an invariant under the finite modular
group is given by
\begin{equation}
\label{kahler_inv}
     (-\I\tau + \I\bar{\tau} )^{k_j} \left(\bar\Phi_j \Phi_j\right)_{\rep{1}}\;,
\end{equation}
where the subindex $\rep{1}$ refers to the trivial $\Gamma_N$ singlet(s) resulting 
from tensoring the superfield $\Phi_j$ with its conjugate.

To complete a supersymmetric model based on modular flavor symmetries, we must specify its
superpotential and K\"ahler potential. In terms of the matter fields $\Phi_{j}$,
the superpotential can be expressed as a polynomial of the form
\begin{equation}
  \mathscr{W}(\Phi) = 
  \sum_{i,j,k} \hat{Y}_{\rep{s}}^{(k_Y)}(\tau)\, \Phi_{i}\Phi_{j}\Phi_{k} 
  + \text{higher order terms}\;,
\end{equation}
where $\hat{Y}_{\boldsymbol{s}}^{(k_Y)}(\tau)$ are modular forms of level
$N$ and modular  weights $k_Y$ transforming as an $s$--dimensional
representation of $\Gamma_N$. In general, the superpotential is constrained to 
transform according to
\begin{equation}
\label{eq:generalW}
 \mathscr{W}(\Phi) \xmapsto{~\gamma~} \mathscr{W}(\Phi') := (c\tau+d)^{k_{\mathscr{W}}} \mathscr{W}(\Phi)\,.
\end{equation}
In our case, given our bottom-up approach, we choose the superpotential  to be
modular invariant, i.e., $ k_{\mathscr{W}}=0$. This amounts to demanding 
$\rep{s}\otimes\rep{r}_i\otimes\rep{r}_j\otimes\rep{r}_k\stackrel{!}{\supset}\rep1$
and $k_Y \overset{!}{=}  -k_i - k_j - k_k$.

The K\"ahler potential of matter fields in models endowed with a modular flavor 
symmetry is typically assumed to take the canonical form
\begin{equation}
\label{eq:generalK}
  K(\Phi,\bar{\Phi}) \supset \sum_j (-\I \tau + \I\bar{\tau})^{k_j}|\Phi^{j}|^2\,,
\end{equation}
as in~\cite{Feruglio:2017spp}. However, the nonlinear realization of this
symmetry implies that there are additional terms with free coefficients,
which are at the same footing as the canonical terms, thus limiting
the predictive power of the model~\cite{Chen:2019ewa}.

\subsection{Eclectic flavor symmetries}
\label{sec:eclectic}

The so--called eclectic flavor symmetries~\cite{Nilles:2020nnc} arise naturally 
in string models~\cite{Nilles:2020kgo,Nilles:2020tdp,Baur:2021mtl} and magnetized 
toroidal compactifications~\cite{Ohki:2020bpo}. They are given by group--theoretic 
unions of a traditional (flavor) symmetry, $G_\mathrm{traditional}$, and a modular
symmetry, $G_\mathrm{modular}$,
\begin{equation}
\label{eq:G_eclectic}
G_\mathrm{eclectic} = G_\mathrm{traditional}\cup G_\mathrm{modular}\;,
\end{equation}
such that the modular symmetry is built out of  outer automorphisms of
$G_\mathrm{traditional}$,  $G_\mathrm{modular} \subset
\mathrm{Out}(G_\mathrm{traditional})$.  The union ``$\cup$'' in
\eqref{eq:G_eclectic} is to be understood as the multiplicative closure of
the groups.

Crucially, $G_\mathrm{eclectic}$ has representations which transform
nontrivially under both $G_\mathrm{traditional}$ and $G_\mathrm{modular}$. This
means that, by giving a \ac{VEV} to a representation of that kind, we can break
$G_\mathrm{eclectic}$ to a diagonal subgroup which inherits properties from
$G_\mathrm{traditional}$ as well as $G_\mathrm{modular}$.  

Even though eclectic groups can also be built from a bottom--up perspective
\cite{Nilles:2020nnc}, in this work we refrain from working out an explicit 
eclectic model. Rather, in what follows we will analyze the somewhat simpler
situation in which the union ``$\cup$'' in \eqref{eq:G_eclectic} gets replaced
by a direct product, i.e.\ $G_\text{quasi-eclectic}=G_\mathrm{traditional}\times
G_\mathrm{modular}$. As we shall see, the emerging scheme is still simple enough
to be analyzed and at the same time illustrates how the desirable properties of
$G_\mathrm{traditional}$ and $G_\mathrm{modular}$ get inherited by the diagonal
group.

\section{A simple quasi--eclectic example}

\subsection{Symmetries and representations}

To illustrate the main points of our quasi--eclectic scheme, let us consider  a
model by Feruglio~\cite{Feruglio:2017spp}, but with a slight twist. We will take
the original flavor symmetry to be
\begin{equation}
 G_\mathrm{flavor}=A_4^\mathrm{traditional}\times\Gamma_3\;,
\end{equation}
where $\Gamma_3$ can be thought of as a modular version of $A_4$. The quantum
numbers of the states are listed in \Cref{tab:TwistedFeruglioModel}. We
take the superpotential to have modular weight $k_{\mathscr{W}}=0$.

\begin{table}[htb]
\centering
$\begin{array}{*{11}{c}}
\toprule
  & (E_1^\ChargeC,E_2^\ChargeC,E_3^\ChargeC) &   L & H_d & H_u & \chi & \varphi  & S_\chi & S_\varphi  
  & Y\\
\midrule 
 \SU{2}_\mathrm{L}\times\U1_Y & \rep{1}_1 &  \rep{2}_{-\nicefrac{1}{2}}
  & \rep{2}_{-\nicefrac{1}{2}} & \rep{2}_{\nicefrac{1}{2}} & \rep{1}_0
  & \rep{1}_0 
  & \rep{1}_0 
  & \rep{1}_0 
  &  \rep{1}_0\\
\midrule 
 A_4^\mathrm{traditional} & (\rep{1}_0,\rep{1_2},\rep{1_1}) &  \rep{3} &
 \rep{1}_0 & \rep{1}_0
 & \rep{3} & \rep{3} 
 & \rep{1}_0 
  & \rep{1}_0 
 & \rep{1}_0\\
\midrule 
 \Z{3}^\chi & 0 & 0 &  0 & 1 & 1 & 0 & 1 & 0 & 0\\
 \Z{3}^\varphi & 1 & 0 &  1 & 0 & 0 & 1 & 0 & 1 & 0\\
\midrule 
 \Gamma_3 & \rep{1}_0 &  \rep{1}_0 & \rep{1}_0 & \rep{1}_0  & \rep{3} & \rep{1}_0 & \rep{1}_0   & \rep{1}_0  & \rep{3} \\
\midrule 
 k & (k_{E_1},k_{E_2},k_{E_3}) &  k_L & k_{H_d} & k_{H_u} & k_\chi & k_\varphi & k_S & k_S & k_Y \\
 \text{modular weights} & (1,1,1) &  -1 & 0 & 0 & 0 & 0 & 0 & 0 & 2 \\
\bottomrule
\end{array}$ %
\caption{Variation of model 1 of \cite{Feruglio:2017spp}. $E_i^\ChargeC$, $L$, $H_u$
and $H_d$ are the superfields of the charged leptons, left--handed doublets,
up--type Higgs and down--type Higgs, respectively. $S_\chi$ and
$S_\varphi$ are part of the \ac{VEV} alignment, see
\Cref{sec:FlavonVEValignment}. In our notation, $A_4\cong\Gamma_3$ has the
representations $\rep3,\rep1_0,\rep1_1$ and $\rep1_2$, whose tensor products
are given e.g.~in~\cite[Appendix~C]{Feruglio:2017spp}.}
\label{tab:TwistedFeruglioModel}
\end{table}

\subsection{Diagonal breaking}

Let us now \emph{assume} that the flavon $\chi$ attains a ``diagonal''
\ac{VEV}, i.e.\ in the real basis 
\begin{equation}\label{eq:phi_diagonal_VEV_real_basis}
 \Braket{\chi^a_i}=v_1\,\mathds{1}_3\;.
\end{equation}
In the complex basis, this diagonal \ac{VEV} has the shape\footnote{The relation
between these bases is explained in \Cref{sec:basisChange}.}
\begin{equation}\label{eq:phi_diagonal_VEV}
 \Braket{\chi^a_i}=v_1\,\begin{pmatrix}
  1 & 0 & 0 \\ 0 & 0 & 1\\ 0 & 1 & 0 
 \end{pmatrix}\;.
\end{equation}
We discuss the alignment of the flavon in \Cref{sec:FlavonVEValignment}.
Similarly to Feruglio, we introduce a flavon $\varphi$ (as in
\cite{Feruglio:2017spp}). Here, $a$ is a $\Gamma_3$ index and $i$ an
$A_4^\mathrm{traditional}$ index. The \ac{VEV} \eqref{eq:phi_diagonal_VEV}
breaks $A_4^\mathrm{traditional}\times\Gamma_3$ to $\Gamma_3^\mathrm{diagonal}$.
Both $\Gamma_3$ and $\Gamma_3^\mathrm{diagonal}$ are nonlinearly realized.

\subsection{Charged lepton Yukawa couplings}
\label{sec:ChargedLeptonYukawaCouplings}

The charged fermion masses are obtained just like in \cite{Feruglio:2017spp}.
Since we assigned them the $\rep{1}_0$, $\rep{1}_1$ and $\rep{1}_2$ under
$A_4^\mathrm{traditional}$, respectively, we can write down superpotential
terms\footnote{Following Feruglio's model (cf.\ \cite[discussion
between Equations (39) and (40)]{Feruglio:2017spp}), we exchange here
$\widetilde{y}_\mu$ and $\widetilde{y}_\tau$ to best fit data.}
\begin{equation}
\label{eq:W_e}
 \mathscr{W}_e=
 \frac{\widetilde{y}_e}{\Lambda}H_d(L\varphi E^\ChargeC_1)_{\rep{1}_0}
 +\frac{\widetilde{y}_\tau}{\Lambda} H_d(L\varphi E^\ChargeC_2)_{\rep{1}_0}
 +\frac{\widetilde{y}_\mu}{\Lambda} H_d(L\varphi E^\ChargeC_3)_{\rep{1}_0}
 \;,
\end{equation}
which involve the three free parameters $\widetilde{y}_e$, $\widetilde{y}_\mu$
and $\widetilde{y}_\tau$, and the cut--off scale $\Lambda$ of the model. Here, 
a $\rep{1}_0$ subscript indicates a contraction to a $G_\mathrm{flavor}$ singlet. 
In order to get a diagonal charged lepton Yukawa coupling matrix, we will take 
the \ac{VEV} of $\varphi$ to be 
\begin{equation}
\label{eq:chi_vev_complex}
\Braket{\varphi_i} = v_{2} \begin{pmatrix}
1 \\ 0 \\ 0 \end{pmatrix}
\end{equation}
in the complex basis and
\begin{equation}
\label{eq:chi_vev_real}
\Braket{\varphi_i} = \frac{v_{2}}{\sqrt{3}} \begin{pmatrix}
1 \\ 1 \\ 1 \end{pmatrix}
\end{equation} 
in the real basis, similarly to Feruglio's model \cite{Feruglio:2017spp}. This
choice will be justified in \Cref{sec:FlavonVEValignment}. \Cref{eq:W_e}
along with \Cref{eq:chi_vev_complex} gives the charged lepton mass matrix
\begin{equation}\label{eq:Me}
m_{e}= v_{d}\,\frac{v_{2}}{\Lambda}\diag\left(\widetilde{y}_e,\widetilde{y}_\tau,\widetilde{y}_\mu\right)\,,
\end{equation}
where $v_d$ is the \ac{VEV} of $H_d$, as usual. Like
in~\cite{Feruglio:2017spp}, we introduced three parameters, $\widetilde{y}_e$,
$\widetilde{y}_\mu$ and $\widetilde{y}_\tau$. These parameters can be used to
reproduce the observed charged lepton masses. In order to reproduce the observed
$\tau$ lepton mass, $\varepsilon_2:=v_2/\Lambda$ cannot become too small. This
sector does not really contain any novel ingredients, nor does it by itself make
nontrivial predictions.

\subsection{Weinberg operator}

Like in Feruglio's model~\cite{Feruglio:2017spp} the new ingredients are in the
Weinberg operator, which emerges from the superpotential couplings
\begin{equation}
 \mathscr{W}_\nu=\frac{1}{\Lambda^2}\left[\left(H_u\cdot L\right)\,\chi\,
 \left(H_u\cdot L\right)\,Y\right]_{\rep{1}_0}\;.
\end{equation}
To construct the couplings at the component level, we first contract 
$Y\chi$ to $\Gamma_3$ singlets. Since $\chi$ consists of three $\Gamma_3$
\rep{3}--plets, we obtain an $A_4^\mathrm{traditional}$ triplet
\begin{equation}
 [(Y\chi)_{(\rep{3},\rep{1}_0)}]_i
 =Y_1\chi^1_i +Y_2\chi^3_i+Y_3\chi^2_i\;,
\end{equation}
where $i$ is an $A_4^\mathrm{traditional}$ index. Here, $(\rep{r},\rep{r'})$
means that the contraction transforms as $(\rep{r},\rep{r'})$ under
$A_4^\mathrm{traditional}\times\Gamma_3$. This $A_4^\mathrm{traditional}$
triplet can be contracted with the unique $A_4^\mathrm{traditional}$ triplet
that emerges from combining the $A_4^\mathrm{traditional}$ triplet $L$ with
itself,
\begin{equation}
 (LL)_{(\rep{3},\rep{1}_0)} =
 \frac{2}{\sqrt{3}}
 \begin{pmatrix}
  L_{1}^2-L_{2} L_{3}\\ L_{3}^2-L_{1} L_{2}\\ L_{2}^2-L_{1} L_{3}
 \end{pmatrix}\;.
\end{equation}
After inserting the ``diagonal'' \ac{VEV}~\eqref{eq:phi_diagonal_VEV},
the effective superpotential coincides, up to an irrelevant prefactor, 
with the one proposed in \cite{Feruglio:2017spp},
\begin{equation}\label{eq:W_nu_eff}
 \mathscr{W}_\nu=\frac{v_1}{\Lambda^2}\left[\left(H_u\cdot L\right)\,
 Y\,
 \left(H_u\cdot L\right)\right]_{\rep{1}_0}\;.
\end{equation}
In particular, the matrix structure of the Weinberg operator is identical to the
one in \cite{Feruglio:2017spp}. That is, the neutrino mass matrix is given by
\begin{equation}\label{eq:mnu_modular}
 m_\nu=\frac{v_u^{2} \varepsilon_1}{\sqrt{3}\Lambda}\begin{pmatrix}
  2Y_1(\tau) & -Y_3(\tau) & -Y_2(\tau) \\
  -Y_3(\tau) & 2Y_2(\tau) & -Y_1(\tau) \\
  -Y_2(\tau) & -Y_1(\tau) & 2Y_3(\tau) 
 \end{pmatrix}\;,
\end{equation}
where $\varepsilon_1=v_1/\Lambda$ and $v_u$ is the \ac{VEV} of $H_u$. Then this
matrix has only three free real parameters:  $\Lambda$, $\re\tau$ and $\im\tau$.

\subsection{Kinetic terms}

Before $\chi$ and $\varphi$ attain \acp{VEV}, the K\"ahler potential of the
charged leptons is diagonal because of the presence of
$A_4^\mathrm{traditional}$. Therefore, the K\"ahler potential is under control.
After the breaking to the diagonal flavor symmetry,
\begin{equation}\label{eq:A_4KahlerL}
 K_L ~=~ L^\dagger\,L+\mathcal{O}(\varepsilon_1^2)+\mathcal{O}(\varepsilon_2^2)\;.
\end{equation}
This is because the corrections to the K\"ahler potential come from terms
involving $\chi$ and $\varphi$. A priori these terms are not known. In this work
we ask how much we can limit the effects of these terms in a bottom--up
approach.

Let us first turn our attention to $\chi$. $\chi$ enters the leptonic
superpotential only through the Weinberg operator. Therefore, we cannot place a
stringent lower bound on the size of $v_1$. 

On the other hand, the magnitude of the \ac{VEV} of $\varphi$, $v_2$, is bounded
from below by the requirement to reproduce a realistic $\tau$ Yukawa coupling,
$y_\tau$. The value $y_\tau$ depends on the Higgs \ac{VEV} ratio $\tan\beta$,
$y_\tau=\sqrt{1+\tan^2\beta}\,m_\tau/v_\mathrm{EW}\sim10^{-2}\,\sqrt{1+\tan^2\beta}$
(at tree level), where $v_\mathrm{EW}$ denotes the electroweak \ac{VEV}. 
In~\cite{Criado:2018thu}, the best fits to data are obtained for small
$\tan\beta$, in which case $y_\tau$ is suppressed, and the lower bound on
$v_2$ is less stringent.

At first glance, one may suspect to find linear contributions to the
K\"ahler metric,
\begin{equation}
\label{eq:kahler_linear}
 K\supset(\varphi L L^{\dagger})_{\rep{1}_0}
 \quad\text{and/or}\quad  
 \bigl(\varphi\, E_{i}^{\ChargeC} (E_{i}^{\ChargeC})^{\dagger}\bigr)_{\rep{1}_0}\;.
\end{equation}
However, the terms \eqref{eq:kahler_linear} are forbidden due to the symmetry
$\Z{3}^{\varphi}$ (cf.\ \Cref{tab:TwistedFeruglioModel}). Thus, the first
nontrivial flavon--dependent contributions to the K\"ahler metric are given by
$(L \varphi)^{\dagger} (L\varphi)$ and $ (\varphi \varphi^{\dagger})(E
E^{\dagger})$, which we will call $\Delta K_{L}$ and $\Delta K_{R}$,
respectively. Let us first focus on the $L$ contribution. Considering the
discrete charges of $L$ and $\varphi$, we identify seven
$A_4^\mathrm{traditional}$  invariant terms from the product $(\rep{3}\otimes
\rep{3}\otimes\rep{3}\otimes\rep{3})$. After inserting on the \ac{VEV} of
$\varphi$ \eqref{eq:chi_vev_complex}, these are reduced to only three
nonvanishing invariant contributions to $\Delta K_{L}$, which are associated
with  three independent coefficients $C_{i}$. The resulting contribution to the
K\"ahler metric, in the complex basis, is
\begin{equation}\label{eq:Delta_KL}
\Delta K_{L} ~=~ \frac{v_{2}^{2}}{3\Lambda^2}
\begin{pmatrix} 
3C_{1} + 4 C_{2}  & 0 & 0 \\
0 & 3C_{1}-2C_{2}+2\sqrt{3}C_{3} & 0 \\
0 & 0 & 3C_{1}-2 C_{2}-2\sqrt{3}C_{3}
\end{pmatrix}\,,
\end{equation}
which can be decomposed as 
\begin{equation}
\label{eq:Delta_K2}
\Delta K_{L} ~=~ 
\varepsilon_{2}^{2} \left(C_{1}\,\mathds{1}_3
+\frac{2C_{2}}{3} \diag\left(2,-1,-1\right)
+\frac{2C_{3}}{\sqrt{3}}
\diag\left(0,1,-1\right)\right)\;.
\end{equation}
In the case of the $R$ contribution, after evaluating in
\eqref{eq:chi_vev_complex} we get nine invariant terms from which only three are
nonvanishing. The resulting contribution, in both complex and real basis, is
\begin{equation}\label{eq:Delta_KR}
\Delta K_{R} ~=~ \varepsilon_{2}^{2} \diag(D_1,D_2,D_3)\;,
\end{equation}
where $D_{i}$ is defined similarly as in \Cref{eq:Delta_KL}.

The impact of these corrections can be estimated using the discussion in
\cite{Chen:2012ha,Chen:2013aya}. We see that the corrections of the mixing
angles come from $\Delta K_{L}$ only. Generically, the solar angle $\theta_{12}$
is the most sensitive angle in a scheme with inverted mass ordering, its
correction gets enhanced by a factor $m_1^2/\Delta m_\mathrm{sol}^2$, which is
about $34$ in the Feruglio model. The
corrections are also proportional to $\varepsilon_2^2=v_2^2/\Lambda^2\gtrsim
y_\tau^2$. Furthermore since the unperturbed theory has diagonal kinetic terms,
the coefficients of the K\"ahler corrections are also not arbitrarily large.
For corrections associated with the coefficient $C_i$ of the K\"ahler
metric $\Delta K_L$ in~\Cref{eq:Delta_K2}, we find
\begin{equation}\label{eq:DeltaTheta_12}
 \Delta\theta_{12}\simeq C_i\,
 \left(\frac{\varepsilon_2}{0.03}\right)^2
 \cdot\begin{cases} 
  \phantom-0\;,& \text{if $i=1$}\;,\\
          -0.05\;,& \text{if $i=2$}\;,\\
  \phantom-0.01\;,& \text{if $i=3$}\;.
 \end{cases}
\end{equation}
While an exact computation of the coefficient $C_i$ would require a UV
completion of the model (cf.\ e.g.\ \cite{Antoniadis:1992pm,Abe:2021uxb}), we 
make the \ac{EFT} assumption that the coefficients are at most of the order
unity. \Cref{eq:DeltaTheta_12} shows that, if the correction is proportional to
the unit matrix, $\theta_{12}$ does not change, as expected. For small
$\tan\beta$, $\varepsilon_2\sim0.03$ is possible, and the K\"ahler corrections
are comparable to the experimental uncertainties. However, for large
$\tan\beta$, the model we discuss here requires additional ingredients to allow
us to make precise predictions. 

Altogether we see that the K\"ahler corrections are controlled by
$\varepsilon_2$, which also governs the charged lepton Yukawa couplings. In this
regard this bottom--up analysis is somewhat reminiscent of \ac{MFV}
\cite{Chivukula:1987py,Buras:2000dm}. We can hence conclude that the
quasi--eclectic  scheme presented here allows us to construct predictive
bottom--up models with modular flavor symmetries.

\section{Summary and Outlook}
\label{sec:SummaryAndOutlook}

\subsection{Summary}

\begin{figure}[t!]
 \centering
 \includegraphics{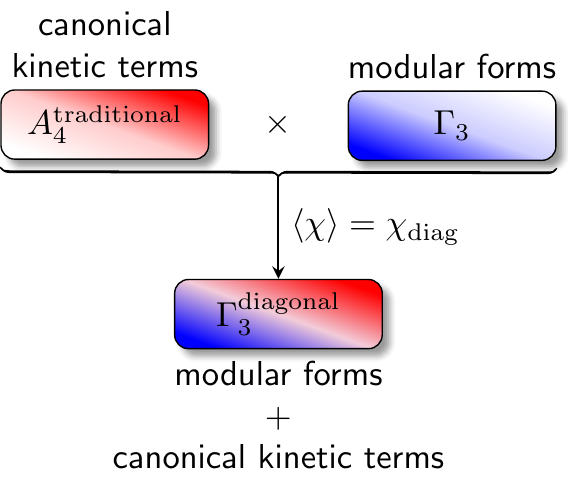}%
\caption{Diagonal breaking of traditional and modular flavor symmetries.}%
\label{fig:DiagonalBreaking}%
\end{figure}

Motivated by the great success of Feruglio's models, we
have proposed a simple way to fix the kinetic terms in this and related 
bottom--up scenarios. To this end, we started with a larger flavor symmetry,
$G_\mathrm{flavor}=A_4^\mathrm{traditional}\times\Gamma_3$, and broke it to its
diagonal subgroup, which is given by the finite modular group $\Gamma_3$ in  our
main example (cf.\ \Cref{fig:DiagonalBreaking}). In the limit of an exact 
$G_\mathrm{flavor}$, the K\"ahler metric is proportional to the unit matrix
because  of $A_4^\mathrm{traditional}$. Therefore, the deviations from canonical
kinetic terms are parametrized by a flavon \ac{VEV}, which also induces the
charged lepton masses, somewhat similarly to the \ac{MFV} scheme. In
particular, we can perform an \ac{EFT} analysis to assess the impact of the
corrections to the K\"ahler potential. We refer to $G_\mathrm{flavor}$ as a
quasi--eclectic modular flavor symmetry, since this bottom--up hybrid scheme 
shares some of the features of top--down  eclectic flavor groups.

We have commented on how to align the \ac{VEV} in such a way that it yields the
desired breaking. Since the corresponding configuration has enhanced symmetries,
it is rather straightforward to achieve this without affecting the lepton
parameters, thus leaving the number of nontrivial predictions unchanged.

The corrections to the K\"ahler potential can still be relevant, but are under
control. Apart from that, they correspond to error bars in our predictions, yet
crucially we are able to specify these error bars, which is, to the best of our
knowledge, not possible if one starts from the nonlinearly realized modular
flavor symmetries alone.

\subsection{Outlook}

We have shown that, by adding ingredients reminiscent to what one finds in
top--down constructions, one can coin predictive bottom--up models with modular
flavor symmetries. Yet it is clear that the toy model presented in this work
leaves some questions unanswered. For instance, we are able to assign the
modular weights and representations at will whereas in top--down models they
derive from the underlying geometry. Also, while the flavon alignment works, it
does not appear to be the final word on this story. One may also envisage flavon
potentials in which the coefficients are modular forms, such that the \acp{VEV}
inherit the pattern from the modular forms. In this case, the kinetic terms may
be under control similarly to what we found in our toy model. It therefore
appears worthwhile to explore similar top--down motivated ingredients to address
the flavor puzzle.

\subsection*{Acknowledgments}

We would like to thank Michael Schmidt for updating the
\href{https://discrete.hepforge.org/}{\texttt{Discrete} package}, and related
correspondence. S.R.-S.\ and M.R.-H.\ thank Alexander Baur for useful
discussions. The work of M.-C.C., M.R.\ and S.S.\ was supported by the National
Science Foundation, under Grant No.\ PHY-1915005.  This material is based upon
work supported by a grant (No.\ CN-20-38) from the University of California
Institute for Mexico and the United States (UC MEXUS) and the Consejo Nacional
de Ciencia y Tecnolog\'ia de M\'exico (Conacyt).

\appendix
\section[Flavon VEV alignment]{Flavon \ac{VEV} alignment\footnote{The
computations of this Section can be checked in the attached supplementary
Mathematica notebook which makes use of
\href{https://discrete.hepforge.org/}{\texttt{Discrete} package}.}}
\label{sec:FlavonVEValignment}

We use the flavon \acp{VEV} \eqref{eq:phi_diagonal_VEV} and
\eqref{eq:chi_vev_complex} (cf.\ \Cref{sec:ChargedLeptonYukawaCouplings}),
which in the so--called real basis (cf.\
\cite[Appendix~C]{Feruglio:2017spp}) are given by
\begin{equation}
 \Braket{\chi^a_i}=v_1\,\begin{pmatrix}
  1 & 0 & 0 \\ 0 & 1 & 0\\ 0 & 0 & 1
 \end{pmatrix}
 \quad\text{and}\quad
 \Braket{\varphi_i}=\frac{v_2}{\sqrt{3}}\,\begin{pmatrix}
  1 \\ 1 \\ 1 \\ 
 \end{pmatrix} 
 \;.\label{eq:our_VEVs}
\end{equation}
\begin{wraptable}[10]{r}[10pt]{6cm}
\centering
$\begin{array}{r*{4}{c}}
\toprule
 \text{field} & \chi & \varphi & S_\chi & S_\varphi\\
\midrule 
 A_4^\mathrm{traditional}  & \rep{3} & \rep{3}  & \rep{1}_0 & \rep{1}_0\\
\midrule 
 \Gamma_3  & \rep{3} & \rep{1}_0  & \rep{1}_0 & \rep{1}_0\\
\midrule 
 \Z3^\chi  & 1 & 0  & 1 & 0\\
\midrule 
 \Z3^\varphi  & 0 & 1  & 0 & 1\\
\bottomrule
\end{array}$ %
\caption{Flavon sector. }
\label{tab:FlavonSector}
\end{wraptable}
We assume that the $A_4^\mathrm{traditional}$ representation matrices act from
the left, and the $\Gamma_3$ matrices act from the right. Then the \ac{VEV}
$\Braket{\chi^a_i}$ is the unique \ac{VEV} which is invariant simultaneous $S$
and $T$ transformations from both groups,
\begin{equation}
 S\,\Braket{\chi}\,S^T=T\,\Braket{\chi}\,T^T=\Braket{\chi}\;.
\end{equation}
So it is a symmetry--enhanced point, which suggests that it should not be too
difficult to obtain such \acp{VEV} \cite{Kofman:2004yc}.

One can make this more explicit. Let us consider the most general 
renormalizable superpotentials involving the flavons $\chi$,
$\varphi$, $S_\chi$ and $S_\varphi$,
\begin{equation}
 \mathscr{W}=\mathscr{W}_\chi+\mathscr{W}_\varphi\;,
\end{equation}
where
\begin{subequations}
\begin{align}
 \mathscr{W}_\chi&=\frac{\kappa_\chi}{2}S_\chi\,(\chi\chi)_{\rep{1}_0}
 -\frac{\lambda_1}{3}(\chi\chi\chi)_{\rep{1}_0}^{(1)}
 -\frac{\lambda_2}{3}(\chi\chi\chi)_{\rep{1}_0}^{(2)}\;,\\
 \mathscr{W}_\varphi&=\frac{\kappa_\varphi}{2}S_\varphi(\varphi\varphi)_{\rep{1}_0}
 -\frac{\lambda_3}{3}(\varphi\varphi\varphi)_{\rep{1}_0}\;.
\end{align}
\end{subequations}
Here, the subscript ``$\rep{1}_0$'' indicates the contraction to a singlet.
There are two independent such contractions of three $\chi$ fields,
\begin{subequations}
\begin{align}
 (\chi\chi\chi)_{\rep{1}_0}^{(1)}&=\chi_{1}^1 \chi_{3}^2 \chi_{2}^3+\chi_{2}^1 \chi_{1}^2
   \chi_{3}^3+\chi_{3}^1 \chi_{2}^2 \chi_{1}^3\;,\\
 (\chi\chi\chi)_{\rep{1}_0}^{(2)}&=\chi_{1}^1 \chi_{2}^2 \chi_{3}^3+\chi_{2}^1 \chi_{3}^2
   \chi_{1}^3+\chi_{3}^1 \chi_{1}^2 \chi_{2}^3  \;.
\end{align}
\end{subequations}
We assume that $S_\chi$ and $S_\varphi$ acquire \acp{VEV}
$\Braket{S_\chi}\ll\Lambda$ and $\Braket{S_\varphi}\ll\Lambda$. This is plausible
since in string--derived models often the \acp{VEV} get fixed by $D$--terms
\cite{Binetruy:1994ru,Binetruy:1996xk}. In fact, in the heterotic orbifold
models, which underlie the eclectic scheme, the Fayet--Iliopoulos (FI)
$D$--terms drive the flavons to nonzero \acp{VEV} \cite{Atick:1987gy}, which has
been verified in many explicitly constructed models (cf.\ e.g.\
\cite{Lebedev:2006kn}). We then denote $\mu_1:=\kappa_\chi\,\Braket{S_\chi}$ and
$\mu_2:=\kappa_\varphi\,\Braket{S_\varphi}$.  One can verify that there is a
nontrivial solution to the $F$--term equations, where the \acp{VEV} are given by
\eqref{eq:our_VEVs} with
\begin{equation}
 v_1=\frac{\mu_1}{\lambda_2}\quad\text{and}\quad
 v_2=\frac{\mu_2}{\lambda_3}
 \;.
\end{equation}
All directions are stabilized. Of course, there is another solution at which all
\acp{VEV} vanish, and there are solutions in which only one of the \acp{VEV}
vanish. Technically, in supergravity the above solution is the deepest minimum
of the scalar potential, but addressing the vacuum energy is beyond the scope of
this study. We also note that at higher orders there are additional terms that
can alter the above solution slightly. Especially cross terms between
$\chi$ and $\varphi$ can shift the \acp{VEV}. However, these terms
appear at much higher order, and are thus suppressed against the K\"ahler
corrections which we discuss and tame in the main text. Altogether we find that,
in a bottom--up \ac{EFT} theory approach we can successfully align the \acp{VEV}
to provide us with a scenario of diagonal breaking
$A_4^\text{traditional}\times\Gamma_3\to\Gamma_3^\text{diagonal}$.

\section{Basis change}
\label{sec:basisChange}

Considering the three-dimensional representation of $A_4$, the group generators 
can be expressed in the \textit{complex basis},
\begin{equation}
\label{eq:realbasis}
 S^C_{\rep{3}} = \frac{1}{3}
 \begin{pmatrix}
  -1 & 2 & 2  \\ 2 & -1 & 2  \\ 2& 2 & -1
 \end{pmatrix}
 \;,\qquad
 T^C_{\rep{3}} =
 \begin{pmatrix}
  1 & 0 & 0 \\ 0 & \omega  & 0 \\ 0 & 0 & \omega ^2
 \end{pmatrix}\;,
\end{equation}
where $\omega = \exp \left(\nicefrac{2\pi\I }{3}\right)$. However, one might find it useful to express 
these generators in the  \emph{real basis}, as we do in \Cref{sec:FlavonVEValignment}, where  they adopt 
the form
\begin{equation}
\label{eq:complexbasis}
 S^R_{\rep{3}} =
\begin{pmatrix}
  1 & 0 & 0  \\ 0 & -1 & 0  \\ 0& 0 & -1
 \end{pmatrix}
 \;,\qquad
 T^R_{\rep{3}} =
 \begin{pmatrix}
  0 & 1 & 0 \\ 0 & 0 & 1 \\ 1 & 0 & 0
 \end{pmatrix}\,.
\end{equation}
These bases are related by the unitary transformation
\begin{equation}
\label{eq:gentransf}
S^{R}_{\rep{3}} = U S^{C}_{\rep{3}} U^{\dagger}
 \quad\text{and}\quad  
T^{R}_{\rep{3}} = U T^{C}_{\rep{3}} U^{\dagger}\;,
\end{equation}
where $U$ is a unitary matrix, given by
\begin{equation}
 U = \frac{1}{\sqrt{3}}
 \begin{pmatrix}
  1 & 1        & 1 \\
  1 & \omega   & \omega^2 \\
  1 & \omega^2 & \omega
\end{pmatrix}\;.
\end{equation}


\providecommand{\bysame}{\leavevmode\hbox to3em{\hrulefill}\thinspace}
\frenchspacing
\newcommand{\origttfamily}{}
\let\origttfamily=\ttfamily
\renewcommand{\ttfamily}{\origttfamily \hyphenchar\font=`\-}

\begin{acronym}
\acro{VEV}{vacuum expectation value}
\acro{EFT}{effective field theory}
\acro{MFV}{minimal flavor violation}
\end{acronym}

\end{document}